\documentclass{PoS}

\usepackage{bbold}
\usepackage{amsmath}

\title{The $D$ to $K$ and $D$ to $\pi$ semileptonic decay form factors from Lattice QCD}

\ShortTitle{The $D$ to $K$ and $D$ to $\pi$ semileptonic decay form factors from Lattice QCD}

\author{The HPQCD Collaboration}

\author{\speaker{Jonna Koponen}\\%         \thanks{}\\
        University of Glasgow\\
        E-mail: \email{jonna.koponen@glasgow.ac.uk}}

\author{Christine T.H. Davies\\
        University of Glasgow}

\author{Gordon Donald\\
        University of Glasgow}

\author{Eduardo Follana\\
        Universidad de Zaragoza}

\author{G. Peter Lepage\\
        Cornell University}

\author{Heechang Na\\
        The Ohio State University}

\author{Junko Shigemitsu\\
        The Ohio State University}

\abstract{
We present a new and very high statistics study of $D$ and $D_s$ semileptonic
decay form factors on the lattice. We work with MILC $N_f=2+1$ lattices and
use the Highly Improved Staggered Action (HISQ) for both the charm and the
light valence quarks. We use both scalar and vector currents to determine the
form factors $f_0(q^2)$ and $f_+(q^2)$ for a range of $D$ and $D_s$ form factors
including those for $D$ to $\pi$ and $D$ to $K$ semileptonic decays. By using
a phased boundary condition we are able to tune accurately to $q^2=0$. We also
compare the shape in $q^2$ to that from experiment. We show that the form
factors are very insensitive to the spectator quark: $D$ to $K$ and $D_s$ to
$\eta_s$ form factors are essentially the same, and the same is true for $D$
to $\pi$ and $D_s$ to $K$. This has important implications when considering
the corresponding $B$/$B_s$ processes.}

\FullConference{XXIX International Symposium on Lattice Field Theory \\
		 July 10 -- 16 2011\\
		 Squaw Valley, Lake Tahoe, California}

\begin{document}

\section{Scalar and vector currents}

\begin{figure}[hb]
\centering
\includegraphics[angle=-90,width=0.3\textwidth]{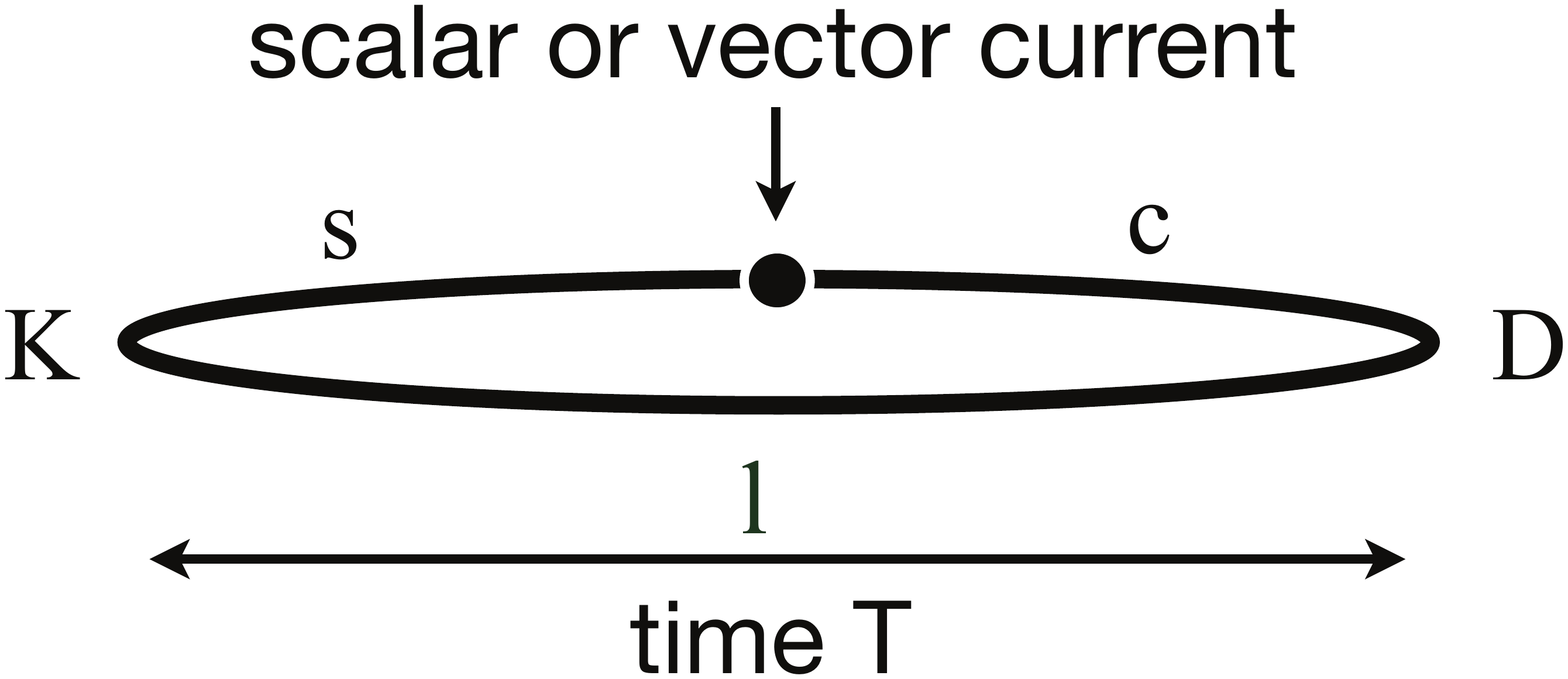}
\caption{Diagram of the 3-point correlator setup.}
\label{fig:3ptdiagram}
\end{figure}

\begin{figure}
\centering
\includegraphics[width=0.55\textwidth]{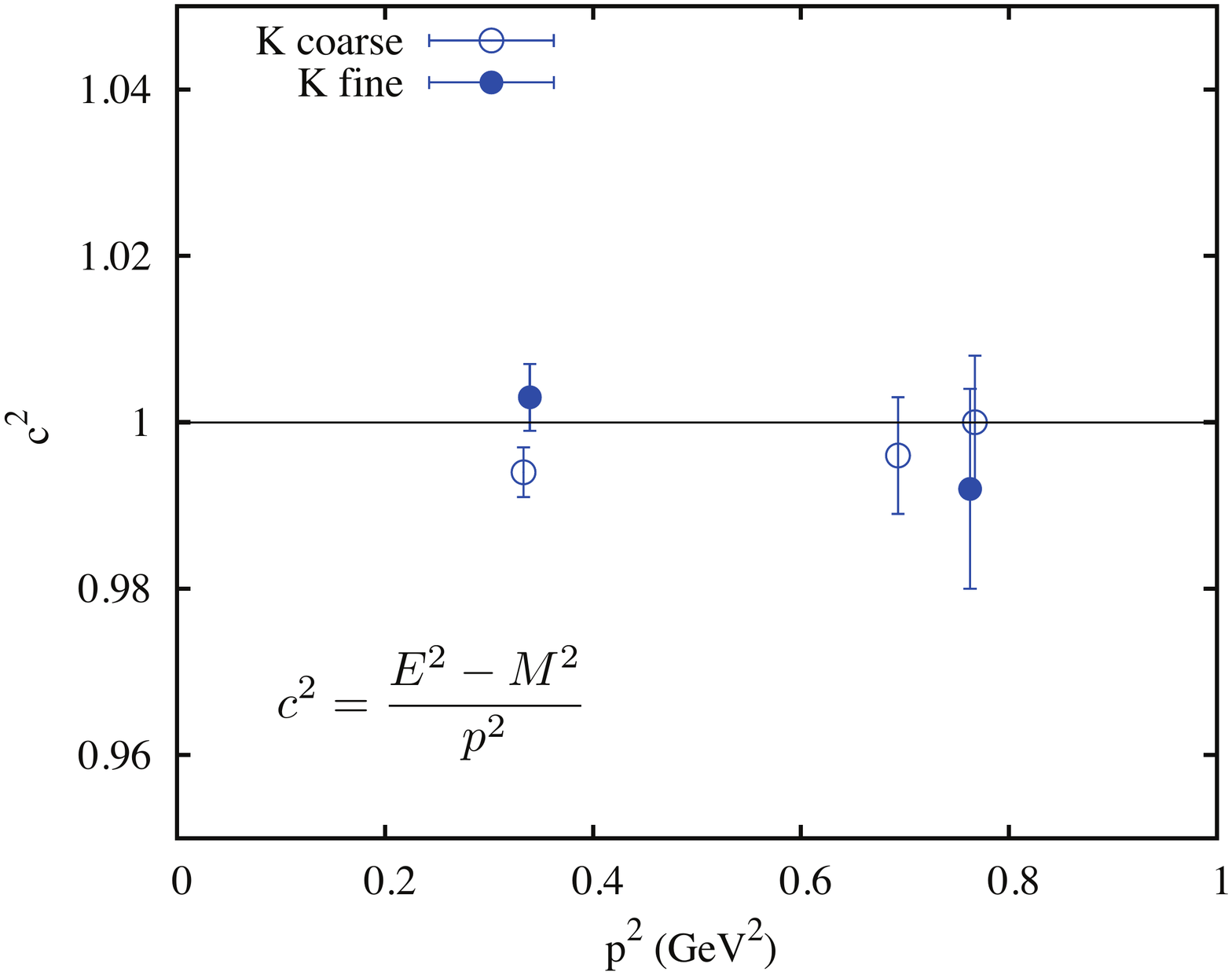}
\caption{Test: Speed of light.}
\label{fig:c2test}
\end{figure}

Form factors $f_0$ and $f_+$ can be extracted from scalar and vector 3-point
correlators --- see diagram in Fig.~\ref{fig:3ptdiagram}. The scalar current
is a local, conserved current 
\begin{equation}
\langle K|S|D\rangle = f^{D\to K}_{0}(q^2)\frac{M^2_D-M^2_K}{m_{0c}-m_{0s}}
\label{equ:f0}
\end{equation}
with $S=\bar{\Psi}\Psi$. Here $q$ is the difference of the four momenta of the mesons,
$q=p_D-p_K$, and the process $D\to Kl\nu$ is used as an example. One or both of
the mesons are given a spatial momentum $p$ using so called twisted boundary
conditions (i.e. a phase at the boundary): 
\begin{equation}
\Phi(x+\hat{e}_jL)=\mathrm{e}^{i2\pi\theta}\Phi(x)
\end{equation}
where $L$ is the size of the lattice. This gives the quark a momentum $2\pi\theta/L$.
Note that $\theta$ can be tuned to get a desired value for $q^2$, e.g. $q^2=0$. We
use different kinematical set-ups: kinematics A, where only one of the mesons has
spatial momentum and the other one is at rest (in Fig.~\ref{fig:3ptdiagram}, $s$ quark
would have the momentum), and kinematics C, where both mesons have the same spatial
momentum (in Fig.~\ref{fig:3ptdiagram}, the light quark would have the momentum).
We have tested this method carefully, for example by checking that the speed of light
is one (see Fig.~\ref{fig:c2test}), and that the amplitude of the meson correlator
depends on the momentum like $1/\sqrt{E}$.

%We have tested this method carefully, for example by checking that the speed of light
%is one, and that the amplitude of the meson correlator depends on the momentum like
%$1/\sqrt{E}$ --- see Figs.~\ref{fig:c2test} and~\ref{fig:Avsptest}.
 
The vector current can be written as 
\begin{equation}
\langle K|V^\mu|D\rangle =
f^{D\to K}_{+}(q^2)\bigg[p^\mu_D+p^\mu_K-\frac{M^2_D-M^2_K}{q^2}q^\mu\bigg]
+f^{D\to K}_{0}(q^2)\frac{M^2_D-M^2_K}{q^2}q^\mu .
\label{equ:fplus}
\end{equation}
We have chosen to use $V^\mu=\gamma^\mu$, a tasteless, spatial vector current. This
has to be a 1-link current, if we have Goldstone mesons. We also tested a local,
temporal vector current $\gamma^t$ with a non-Goldstone $D_s$ ($\gamma_5\gamma_t$)
for $D_s\to\eta_s$ \cite{Latt2011:Gordon} (denoted as $V_t$ in Fig.~\ref{fig:DK}).
We use MILC $N_f=2+1$ lattices to do the calculations --- see Table~\ref{table:configs}
for more details.

%\begin{figure}
%\centering
%\includegraphics[width=0.52\textwidth]{a-k.pdf}
%\caption{Test: Momentum dependence of the amplitude.}
%\label{fig:Avsptest}
%\end{figure}

\begin{table}[bh]
\centering
\begin{tabular}{|c|c|c|c|c|c|}
\hline
ensemble & size, $L^3\times N_t$ & physical size & \# configs. & \# time sources & $m_l$ \\
\hline
coarse & $20^3\times 64$ & $\approx (2.4 \textrm{ fm})^3$ & 2259 & 8 & $\approx m_s/3.5$ \\
fine   & $28^3\times 96$ & $\approx (2.4 \textrm{ fm})^3$ & 1911 & 4 & $\approx m_s/4.2$ \\
\hline
\end{tabular}
\caption{Details of the MILC 2+1 flavor lattice configurations used in this study:
lattice size, number of configurations, number of time sources per configuration,
and light valence quark mass $m_l$ (compared to strange quark mass $m_s$). The mass
values for HISQ valence light quarks are tuned to match the same goldstone pion 
mass as those for the asqtad sea light quarks. The HISQ valence s quark masses are
tuned  to the physical value --- see~\cite{HPQCD_mass} for more details.}
\label{table:configs}
\end{table}

\subsection{Fitting}

To extract the form factors we do a simultaneous least $\chi^2$ fit to both 3-point
correlators and the corresponding 2-point meson correlators. For a given semileptonic
decay, say $D\to K$, we also fit all $q^2$ values simultaneously. For the 2-point
correlators the fit function is the usual sum of exponentials (with the oscillating
states, as we are dealing with staggered quarks): e.g. for the $D$ meson
\begin{equation}
C_D(t)=\sum_j(b_j^D)^2\big(\textrm{e}^{-E^D_jt}+\textrm{e}^{-E^D_j(N_t-t)}\big)-
\sum_k(d_k^D)^2(-1)^t\big(\textrm{e}^{-E'^D_kt}+\textrm{e}^{-E'^D_k(N_t-t)}\big).
\end{equation}
The fit functions for the 3-point correlators have similar form: e.g. for $D\to K$
we have
\begin{align}
C_{D\to K}(t,T)=&\sum_j\sum_kA^{D\to K}_{jk}\textrm{e}^{-E^K_jt}\textrm{e}^{-E^D_k(T-t)}-
\sum_j\sum_kB^{D\to K}_{jk}\textrm{e}^{-E^K_jt}\textrm{e}^{-E'^D_k(T-t)}(-1)^{T-t}\\
-&\sum_j\sum_kC^{D\to K}_{jk}\textrm{e}^{-E'^K_jt}\textrm{e}^{-E^D_k(T-t)}(-1)^{t}+
\sum_j\sum_kD^{D\to K}_{jk}\textrm{e}^{-E'^K_jt}\textrm{e}^{-E'^D_k(T-t)}(-1)^{T},
\end{align}
where $A_{00}=b^K_0b^D_0\langle K|S|D\rangle/(2\sqrt{M_DE_K})$ gives the desired form
factor $f_0$ at a given $q^2$. We use three or four time separations $T$ for the
mesons (see Fig.~\ref{fig:3ptdiagram}), as the 3-point correlators are oscillating.

\section{Renormalization of the currents}

\begin{figure}
\centering
\includegraphics[width=0.55\textwidth]{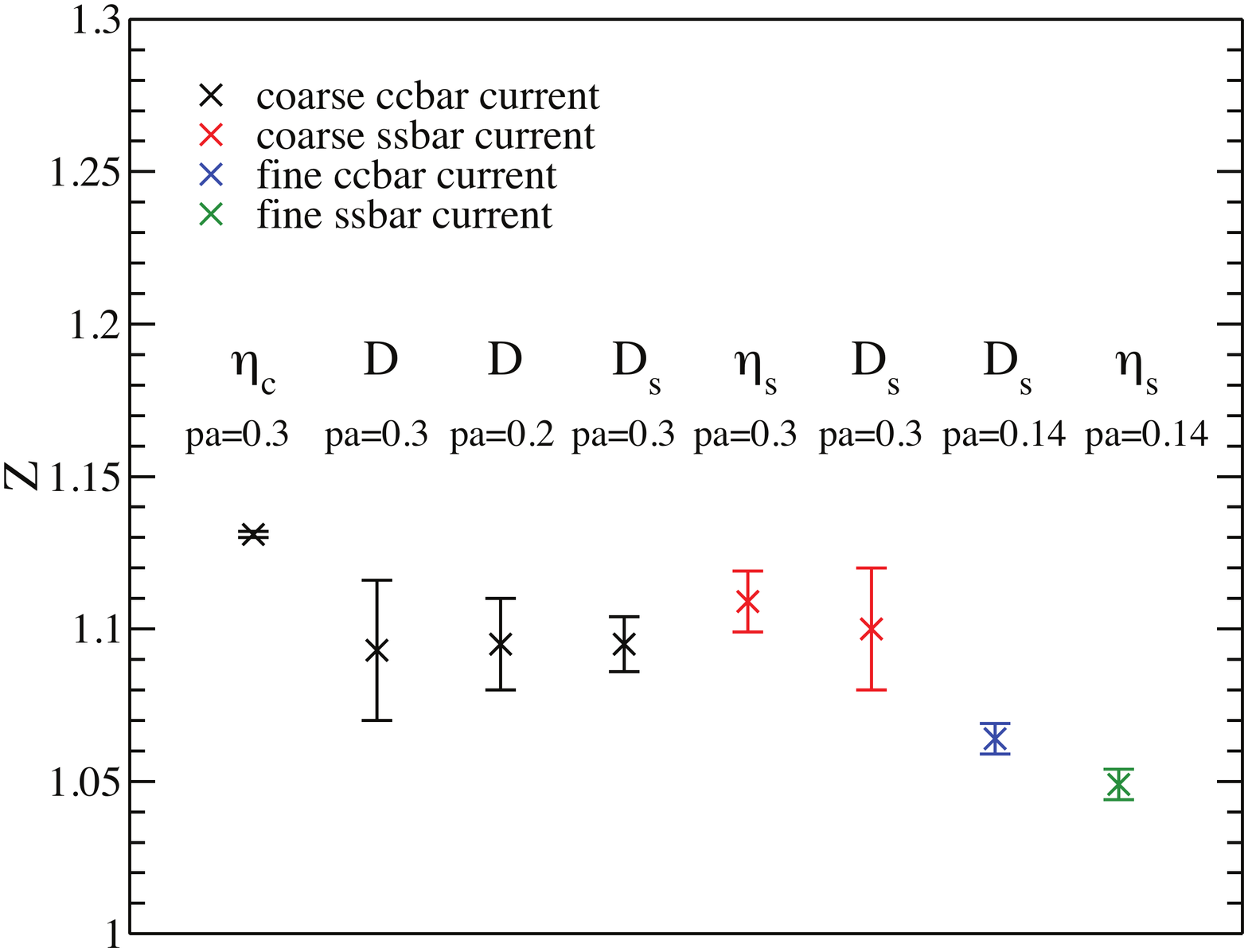}
\caption{Renormalization factor Z for the 1-link vector current.}
\label{fig:vectorZ}
\end{figure}

The scalar current is absolutely normalized (note the bare quark masses in
Eq.~\eqref{equ:f0} --- the renormalization factors cancel). However, the vector
current does need to be renormalized. We extract the renormalization factor
$Z$ from the symmetric vector current by demanding $f_+^{H\to H}(0)=1$ for $H=D$,
$D_s$, $\eta_s$, and $\eta_c$. The extracted $Z$ factors agree over a range of
momenta and different mesons for  both charm-charm and charm-strange currents
--- see Fig.~\ref{fig:vectorZ}. This is essential, as we want to use $Z$ to
renormalize a charm-strange and a charm-light current.

The local, temporal current $V_t$ is renormalized using  $f_0$ at $q^2_\textrm{max}$
extracted from the scalar current. Note that at $q^2_\textrm{max}$ the temporal
vector current gives the form factor $f_0$ directly, as the coefficient multiplying
$f_+$ vanishes (see Eq.~\ref{equ:fplus}). We can thus calculate $f_0(q^2_\textrm{max})$,
and set $f_{0,V_t}(q^2_\textrm{max})=f_{0,S}(q^2_\textrm{max})$.

\begin{figure}
\centering
\includegraphics[width=0.5\textwidth]{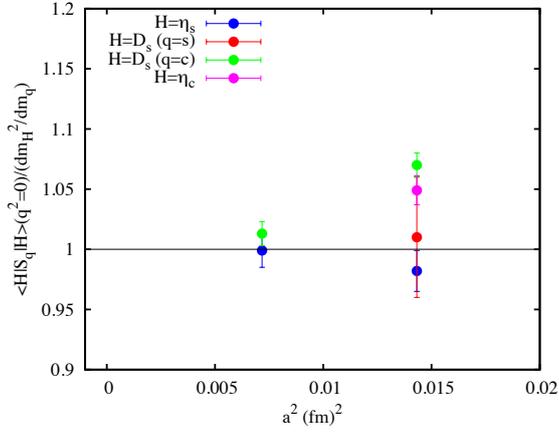}
\caption{Test of the symmetric scalar current.}
\label{fig:symmscalar}
\end{figure}

As yet another test we calculated the symmetric scalar current,
$\langle H|S|H\rangle$, for $H=D_s$, $\eta_s$, and $\eta_c$. At $q^2=0$, and when
lattice spacing $a\to 0$, this is expected to give
\begin{equation}
\langle H|S_q|H\rangle(q^2=0)=\frac{dm_H^2}{dm_q}.
\end{equation}
This is indeed the case, as can be seen in Fig.~\ref{fig:symmscalar}.

\section{Preliminary results: form factors $f_0$ and $f_+$}

\begin{figure}
\centering
\includegraphics[width=0.55\textwidth]{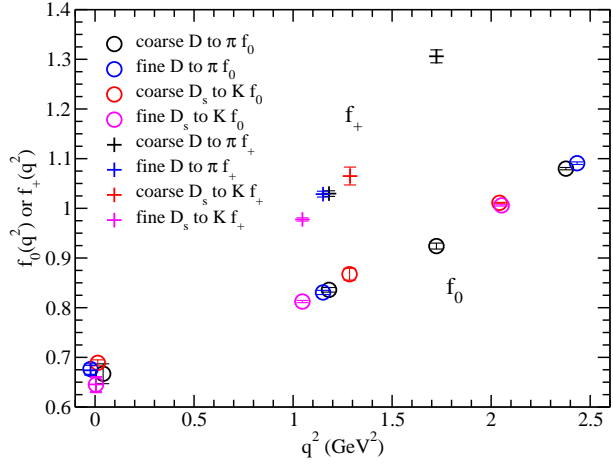}
\caption{Preliminary results: form factors, charm to light decay.}
\label{fig:Dpi}
\end{figure}

\begin{figure}
\centering
\includegraphics[width=0.58\textwidth]{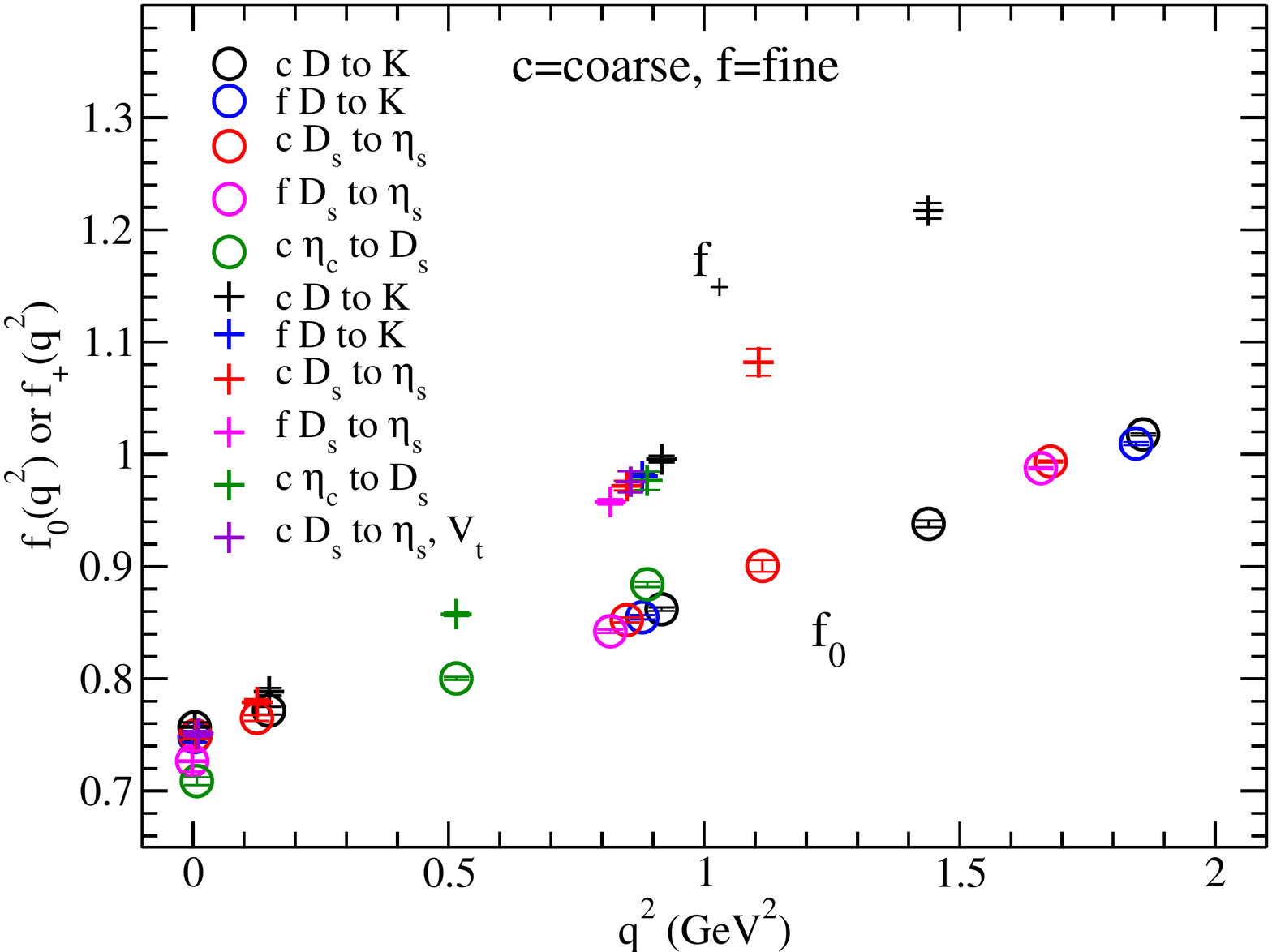}
\caption{Preliminary results: form factors, charm to strange decay.}
\label{fig:DK}
\end{figure}

Our preliminary results for the $D\to \pi$ and $D\to K$ semileptonic decay
form factors $f_0$ and $f_+$ are presented in Figs.~\ref{fig:Dpi},~\ref{fig:DK},
and compared to experimental results in Section~\ref{ExpmntlComp}. We consider
different mesons: For example, we calculate the charm-strange current
3-point correlator with different spectator quarks, light, strange and
charm. This allows us to compare the form factors for $D\to K$,
$D_s\to \eta_s$ and $\eta_c\to D_s$. The semileptonic decay form factors
are very insensitive to the spectator quark: $D\to K$ and $D_s\to \eta_s$
form factors are almost identical. However, if the spectator quark is as heavy
as the charm quark, as in $\eta_c\to D_s$, the form factors do have a noticeably
different shape. The insensitiveness of the form factors to the spectator
quark is also seen in the light-charm current: $D\to \pi$ and $D_s\to K$
form factors have the same shape. One would expect to see similar behaviour in
the corresponding $B/B_s$ form factors, i.e. basically no dependence on the
spectator quark.

\subsection{$z$-expansion and continuum extrapolation}

\begin{figure}
\centering
\includegraphics[width=0.58\textwidth]{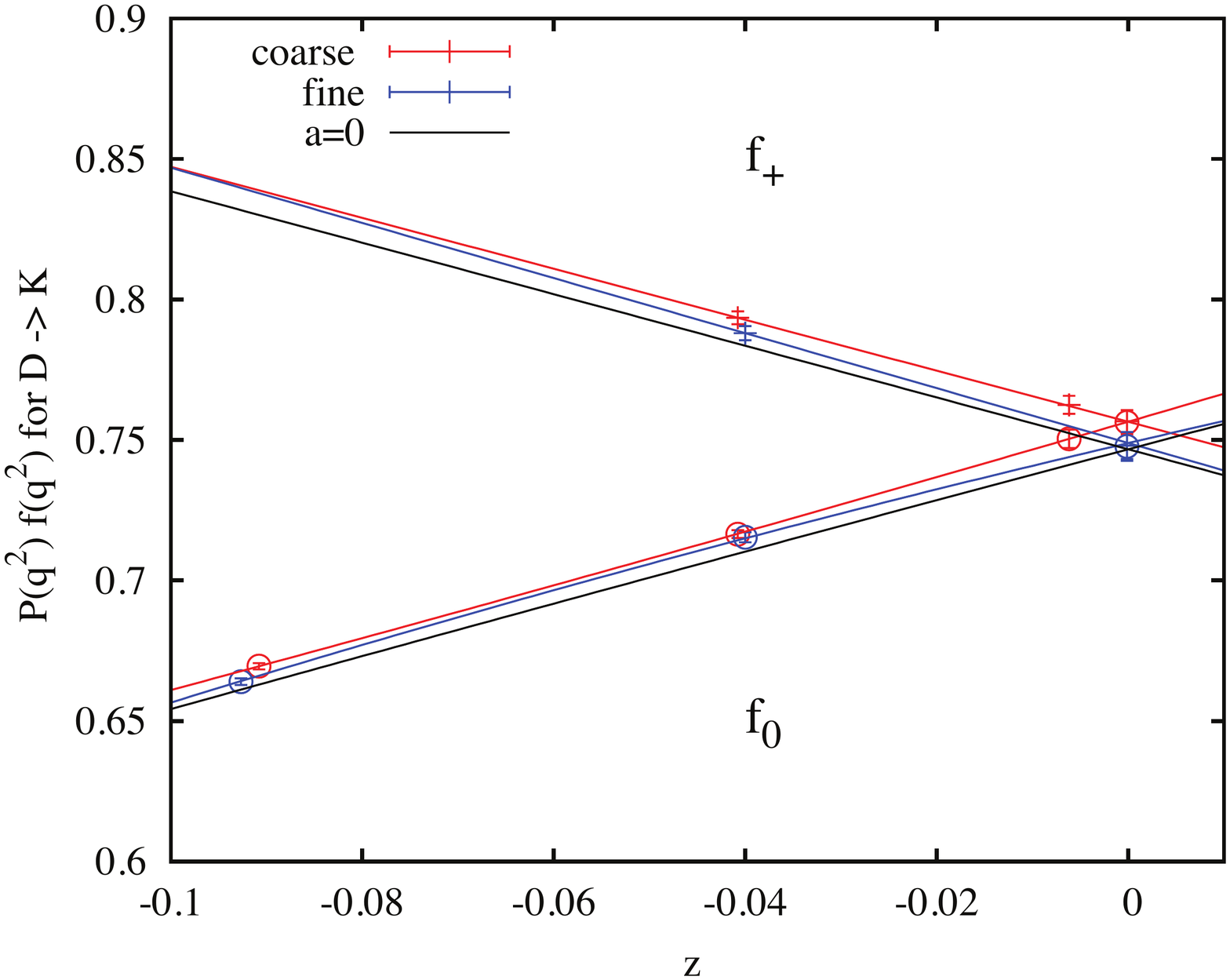}
\caption{Form factors in z-space.}
\label{fig:zspace}
\end{figure}

It is convenient to transform the form factors to $z$-space to do the continuum
extrapolation. This is done as follows: First we remove the poles from the form
factors,
\begin{equation}
\tilde{f}^{D\to K}_0(q^2)=\bigg(1-\frac{q^2}{M^2_{D_{s0}^\ast}}\bigg)f^{D\to K}_0(q^2),\quad
\tilde{f}^{D\to K}_+(q^2)=\bigg(1-\frac{q^2}{M^2_{D_{s}^\ast}}\bigg)f^{D\to K}_+(q^2).
\end{equation}
Here $D$ to $K$ is used as an example, so the poles are $M^2_{D_{s0}^\ast}$ and
$M^2_{D_{s}^\ast}$. We then change from q to z,
\begin{equation}
z=\frac{\sqrt{t_+-q^2}-\sqrt{t_+}}{\sqrt{t_+-q^2}+\sqrt{t_+}},\quad t_+=(m_D+m_K)^2
\end{equation}
--- note that we have taken $t_0=0$ in the standard transformation formula ---
and fit the lattice data as power series in $z$,
\begin{equation}
\tilde{f}^{D\to K}_0(z)=\sum_{n\geq 0}b_n(a)z^n,\quad \tilde{f}^{D\to K}_+(z)
=\sum_{n\geq 0}c_n(a)z^n.
\end{equation}
We include terms up to $z^4$. Note that $b_0=c_0$, because $f_0(q^2=0)=f_+(q^2=0)$.
The lattice spacing dependence is very small, and the extrapolation to $a=0$ is
shown in Fig.~\ref{fig:zspace}.

\subsection{Comparison with experimental results}
\label{ExpmntlComp}

\begin{figure}
\centering
\includegraphics[width=0.48\textwidth]{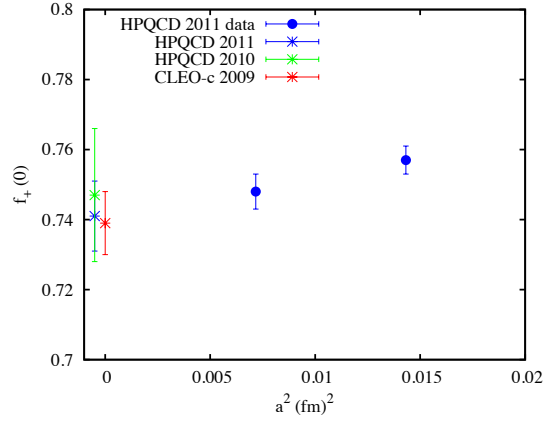}
\caption{Form factor at $q^2=0$, extrapolated to $a=0$.}
\label{fig:fat0cont}
\end{figure}

\begin{figure}
\centering
\includegraphics[width=0.62\textwidth]{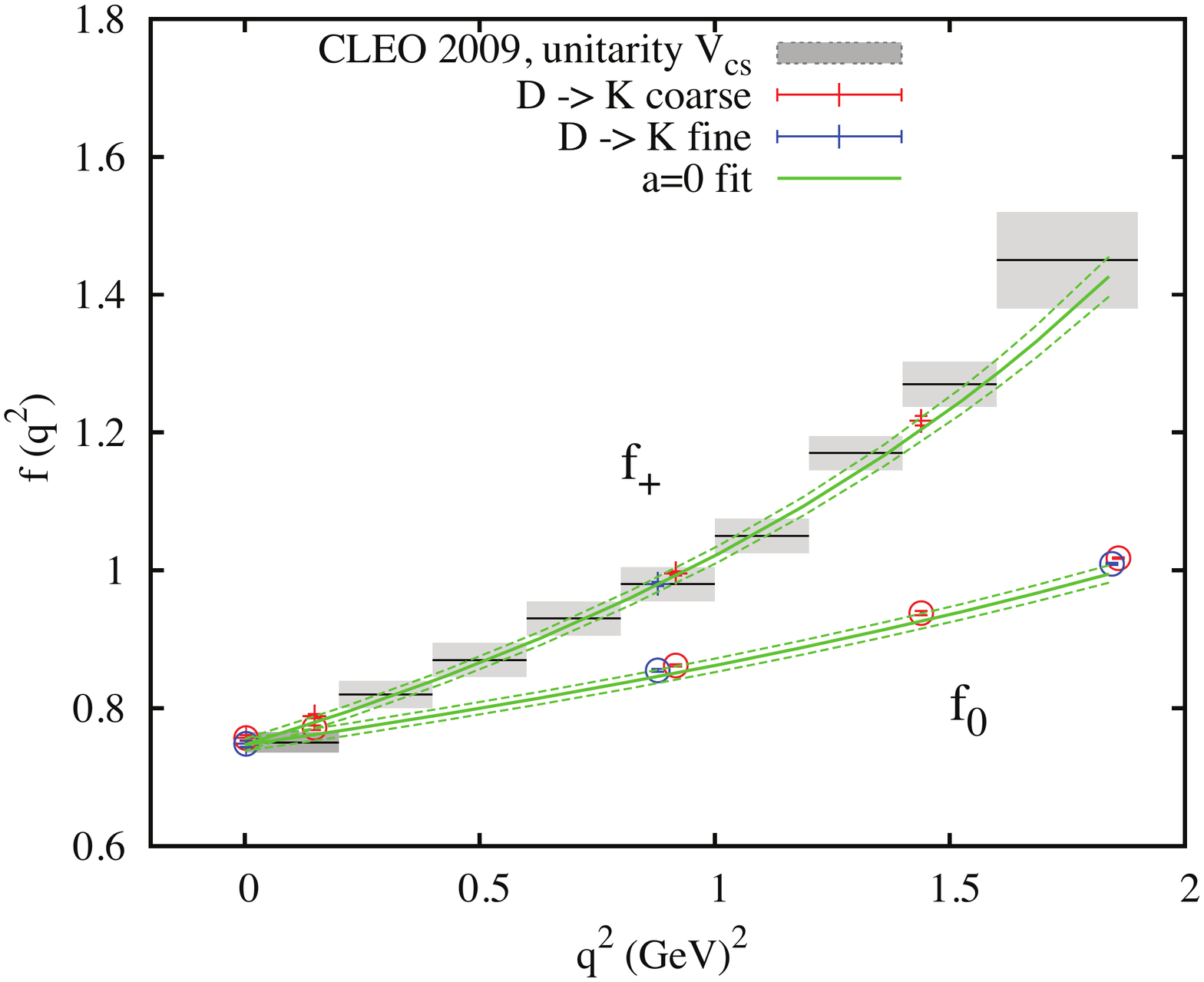}
\caption{$D\to K$ form factors and experimental results from CLEO.}
\label{fig:DKwExperimental}
\end{figure}

In Figs.~\ref{fig:fat0cont} and~\ref{fig:DKwExperimental} we compare our
new results to earlier results as well as experimental results. Earlier
results for $D\to K$ form factor at $q^2=0$ from HPQCD Collaboration are
from~\cite{HPQCD}, and the experimental results
by CLEO Collaboration are from~\cite{CLEO}. Both the value at $q^2=0$ and
the shape of the $D\to K$ form factors $f_0$ and $f_+$ agree very well with
experimental results, where $|V_{cs}|$ is calculated from the CKM matrix
assuming unitarity.

\section{Conclusions}

We have presented here the first results from a very high precision
study of $D$ meson semileptonic decay vector form factors. We have
looked at semileptonic decays of different mesons, and have shown
that the form factors are very insensitive to the spectator quarks ---
we expect the same to be true for $B$/$B_s$ processes.
Indeed our results indicate that $B$ and $B_s$ semileptonic form factors
at a given $q^2$ should differ by less than 5\%, whereas their decay
constants differ by approximately 20\%. QCD sum rules~\cite{QCDsumrules}
predict SU(3) breaking effects at around 10\% for both. We need to repeat
our calculation with a lighter sea light quark mass to do a chiral
extrapolation, and the simulations are already underway.\\

The authors wish to thank the MILC Collaboration for providing the lattice
configurations, the High Performance Computing Centre in Cambridge
(part of the DiRAC facility), and STFC.


\begin{thebibliography}{99}

  \bibitem{Latt2011:Gordon} The HPQCD Collaboration, G. Donald, C.T.H. Davies,
and J. Koponen, PoS(Lattice 2011)278

  \bibitem{HPQCD_mass} C. McNeile, C.T.H. Davies, E. Follana, K. Hornbostel, and
G.P. Lepage, Phys.Rev.D82:034512,2010

  \bibitem{HPQCD} The HPQCD Collaboration, H. Na, C.T.H. Davies, E. Follana,
G.P. Lepage, and J. Shigemitsu, Phys.Rev.D82:114506,2010 and PoS(Lattice 2010)315;
The HPQCD Collaboration, arXiv:1109.1501

  \bibitem{CLEO} The CLEO Collaboration: D. Besson, et al, Phys.Rev.D80:032005,2009

  \bibitem{QCDsumrules} P. Blasi, P. Colangelo, G. Nardulli, and N. Paver,
Phys.Rev.D49:238,1994

\end{thebibliography}
\end{document}